\documentclass[showpacs,aps,prl,floatfix,superscriptaddress]{revtex4}
\usepackage{graphicx}
\usepackage{citesort}
\usepackage{epic,eepic}
\usepackage{graphics}
\usepackage{epsfig}
\usepackage{bm}
\usepackage{longtable}
\usepackage{amssymb}

\usepackage{amsmath}

\newcommand{\be}{\begin{equation}}
\newcommand{\ee}{\end{equation}}
\newcommand{\bea}{\begin{eqnarray}}
\newcommand{\eea}{\end{eqnarray}}
\begin{document}                                                                                   
\title{Linear shear flow past a hemispherical droplet adhering to a solid surface} 
\author{K. Sugiyama} \affiliation{Department of Applied Physics, University of Twente, P.O. Box 217, 7500 AE Enschede, The Netherlands} 
\author{M. Sbragaglia} \affiliation{Department of Applied Physics, University of Twente, P.O. Box 217, 7500 AE Enschede, The Netherlands}

\begin{abstract}
This paper investigates the properties of a three dimensional shear flow overpassing a hemispherical droplet resting on a plane wall. The exact solution is computed as a function of the viscosity ratio between the droplet and the surrounding fluid and generalizes the solution for the hemispherical no-slip  bump given in an earlier paper by Price (1985). Several expressions including the torque and the force acting on the drop will be considered as well as the importance of the deformations on the surface for small Capillary numbers.
\end{abstract}

\maketitle

\section{Introduction}

A detailed description of shear flows past viscous droplets resting on a plane wall is necessary for understanding and modelling a series of engineering, physiological and natural situations \cite{Degennes}. Examples include bubble growth and detachment from heated walls \cite{MacCann,Collin}, drop removal and dislocation in detergency problems \cite{Chatter} and also membrane emulsification, where a liquid is pumped into another, shearing droplets and consequently forming emulsions. Pertinent physiological applications can also be included such as the growth of thrombosis and  also a tentative description of the deformation and dislodging of white and red blood cells adhering to the endothelium.\\
Shear flows over protuberances have already been considered in previous works and studies: O'Neill \cite{Oneill} derived an infinite series solution for the flow over a full sphere in contact with a wall while Hyman \cite{Hyman} and Price \cite{Price} derived a  solution for the case of hemispherical no-slip bumps, a problem considered as a limiting case for highly viscous droplets. Pozrikidis \cite{Pozrikidis} provided an accurate description of shear flows over protuberances projecting from a plane surface using the boundary integral method. A considerable number of papers have also described drop displacement problems \cite{Dussan1,Dussan2,Dussan3,Dimi1,Dimi2}  and their variations: Yon, Li and Pozrikidis \cite{Li,Yon} studied shear induced deformations of droplets with fixed contact lines and the work of Schleizer \cite{Bonnecaze} faced the problem in both pressure driven and shear flows. More recently also inertial effects past pinned droplets have been considered \cite{Spelt}.\\
In this paper we derive an exact solution for the case of an
incompressible shear flow overpassing a hemispherical droplet of arbitrary
viscosity. This solution will be computed as a function of the
viscosity ratio between the droplet and the fluid and is the natural
generalization of the one proposed by Price \cite{Price}, here recovered in
the limit of highly viscous droplet. Several analytical
expressions  including the force and torque acting on the droplet will
be given. Moreover, in the limit of small Capillary numbers,  the role of surface deformations will be investigated and comparison with available numerical data will be presented.  

\section{Formulation of the problem}

\begin{figure}
\begin{center}
\includegraphics[width=.9\textwidth,angle=0]{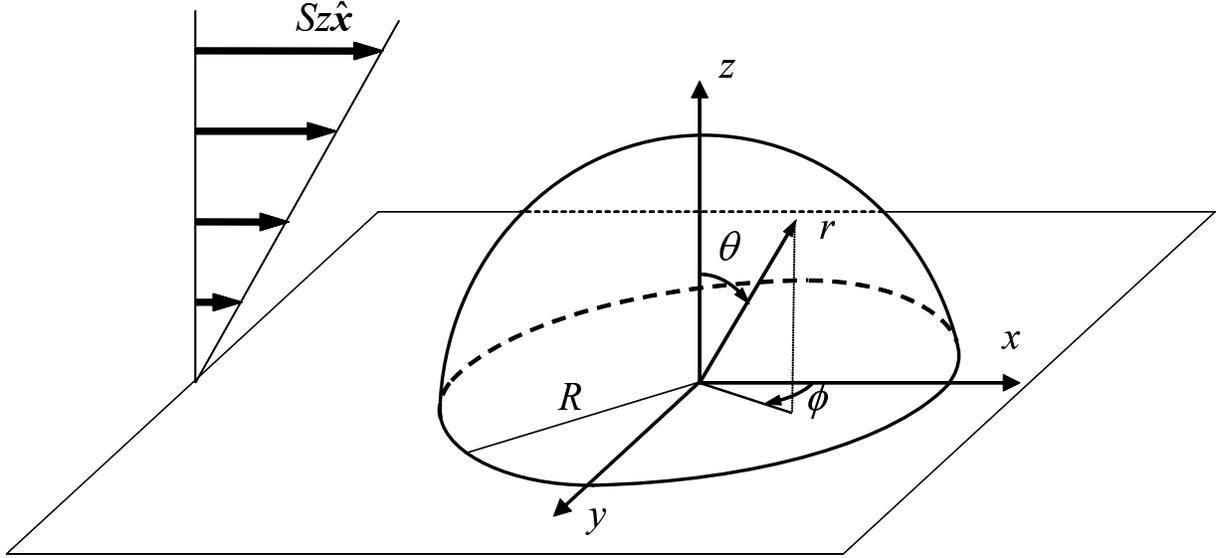}
\caption{Configuration for the problem}
\label{fig:1}
\end{center}
\end{figure}

We consider the motion of a viscous incompressible fluid with velocity ${\bf v}$, of constant density $\rho$ and dynamical viscosity $\eta$, close to a solid plane boundary. Under the assumption of small Reynolds numbers, this motion is reasonably approximated by a uniform shear flow whose magnitude will be denoted with $S$. We will consider how this flow is disturbed by a hemispherical drop of radius $R$ and density $\hat{\rho}$ on the plane boundary. We will use polar coordinates $(r,\theta,\phi)$ with the origin in the center of the hemisphere and an $x$-axis in the direction of the shear flow (see figure \ref{fig:1} for details). The flow inside the drop will be identified with a velocity $\hat{\bf v}$ and viscosity $\hat{\eta}$ in a prescribed  ratio with respect to the outer viscosity. This ratio will be denoted by $\lambda=\hat{\eta}/\eta$. Indicating with $p$ and $\hat{p}$ the outer and inner pressure respectively, the continuity and Stokes equations read as follows:
\begin{equation}
\nabla\cdot{\bf v}=\nabla\cdot\hat{{\bf v}}=0,
\label{eq:ge_cont02}
\end{equation}
\begin{equation}
\nabla\cdot{\bm \sigma}=\nabla\cdot\hat{{\bm \sigma}}=0,
\label{eq:ge_stokes02}
\end{equation}
where 
\begin{equation}\label{eq:Stokes}
{\bm \sigma}= -{p}{\bf I}
+\eta \left(\nabla {\bf v}+(\nabla {\bf v})^T\right),
\ \ \ 
\hat{{\bm \sigma}}=-\hat{{p}}{\bf I}
+\hat{\eta} \left( \nabla\hat{{\bf v}}+(\nabla\hat{{\bf v}})^T \right).
\end{equation}
We will assume no-slip boundary conditions on the wall:
\be
{\bf v}=0 \hspace{.2in} r \ge R,\ \theta=\frac{\pi}{2},\ 0 \le \phi \le 2 \pi,
\ee
\be
\hat{\bf v}=0 \hspace{.2in} r \le R,\ \theta=\frac{\pi}{2},\ 0 \le \phi \le 2 \pi,
\ee
and, on the interface of the hemispherical bump ($r=R$,$0 \le \theta \le \frac{\pi}{2}$,$0 \le \phi \le 2 \pi$), 
we should match the inner and outer solutions by imposing (i) the kinetic condition for two phases, 
(ii) the continuity of the tangential velocity in two directions
and (iii) the continuity of the shearing surface force per unit area in two directions.
These boundary conditions are written in a general form:
\begin{equation}
\begin{split}
{\bm e}_r\cdot {{\bf v}}=&{\bm e}_r\cdot \hat{{\bf v}}=0, \hspace{.2in} \\
{\bm e}_r\times {{\bf v}}=&
{\bm e}_r\times \hat{{\bf v}},\\
{\bm e}_r\times 
\left\{ {\bm \sigma}\cdot {\bm e}_r\right\}
=&
{\bm e}_r\times 
\left\{\hat{{\bm \sigma}}\cdot {\bm e}_r\right\},
\label{eq:bc01}
\end{split}
\end{equation}
with ${\bm e}_r$ the radial normal unit vector at the bump surface.

\section{General Solution}

In the subsequent developments we will assume that all variables have been non-dimensionalized using $S,\rho,\eta$ in the outer region and  $S,\hat{\rho},\hat{\eta}$ in the inner region together with a reference length scale given by the radius of the drop $R$.  We will also introduce the disturbances of the velocity vectors ${\bf v}$ and $\hat{\bf v}$  from the linear shear flow as
\begin{equation}
\begin{split}
{\bm q}=&{\bm e}_xz-{\bf v},
\ \ \ 
\hat{{\bm q}}={\bm e}_xz-\hat{{\bf v}}.\ \ \ 
\label{eq:dist_vel01}
\end{split}
\end{equation}
These disturbances satisfy the continuity and Stokes equations:
\begin{equation}\label{CONTQ}
\nabla\cdot{\bm q}=\nabla\cdot\hat{{\bm q}}=0,
\end{equation}
\begin{equation}\label{STOKESQ}
\nabla^{2} {\bm q}=-{\bm \nabla}p\hspace{.2in}
\nabla^{2} \hat{\bm q}=-{\bm \nabla}\hat{p}.
\end{equation}
The boundary conditions for non slip walls are now expressed by
\be\label{noslip}
{\bm q}=0 \hspace{.2in} r \ge 1,\ \theta=\frac{\pi}{2},\ 0 \le \phi \le 2 \pi,
\ee
\be\label{noslipb}
\hat{\bm q}=0 \hspace{.2in} r \le 1,\ \theta=\frac{\pi}{2},\ 0 \le \phi
\le 2 \pi,
\ee
while, considering the shear velocity ${\bm e}_x z$ written in spherical coordinates, we can  rewrite the boundary conditions (\ref{eq:bc01}) using the disturbance velocities as
\begin{equation}
\begin{split}
q_r-\cos\theta\sin\theta\cos\phi=0,\ \ \ &
\hat{q}_r-\cos\theta\sin\theta\cos\phi=0,\\
q_\theta=\hat{q}_\theta,\ \ \ &
q_\phi=\hat{q}_\phi,\\
\frac{\partial q_\theta}{\partial r}-q_\theta=\lambda\left(
\frac{\partial \hat{q}_\theta}{\partial r}-\hat{q}_\theta
\right),\ \ &
\frac{\partial q_\phi}{\partial r}-q_\phi
=\lambda\left(
\frac{\partial \hat{q}_\phi}{\partial r}-\hat{q}_\phi
\right) \\\hspace{.3in} 
r=1,\ &0 \le \theta \le \pi/2,\  0 \le \phi \le 2 \pi.
\label{eq:bc02}
\end{split}
\end{equation}
To reduce the $\phi$-dependence of the boundary conditions (\ref{eq:bc02}) 
and  obtain an axisymmetrical formulation of the problem, 
we introduce $u$, $v$, $w$, $\hat{u}$, $\hat{v}$ and $\hat{w}$ given by
\begin{equation}
\begin{split}
u=&\frac{q_r}{\cos\phi},\ \ \ 
v=\frac{q_\theta}{\cos\phi},\ \ \ 
w=\frac{q_\phi}{\sin\phi},\nonumber\\
\hat{u}=&\frac{\hat{q}_r}{\cos\phi},\ \ \ 
\hat{v}=\frac{\hat{q}_\theta}{\cos\phi},\ \ \ 
\hat{w}=\frac{\hat{q}_\phi}{\sin\phi}.
\nonumber\\
\end{split}
\end{equation}
We further introduce $U$, $V$ and $W$ as functions of 
the velocity components $u$, $v$ and $w$: 
\begin{equation}
\begin{split}
U=&u\sin\theta+v\cos\theta+w,\ \ \ 
V=u\sin\theta+v\cos\theta-w,\\
W=&u\cos\theta-v\sin\theta,\nonumber\\
\end{split}
\end{equation}
and, for the inner region, $\hat{U}$, $\hat{V}$ and $\hat{W}$ are similarly defined. Then, the boundary conditions on the droplet surface can be rewritten using  $U$, $V$, $W$, $\hat{U}$, $\hat{V}$ and $\hat{W}$ as
\begin{equation}
W\cos\theta+\left(\frac{U+V}{2}\right)\sin\theta
-\cos\theta\sin\theta=0,
\label{eq:kinconou01}
\end{equation}
\begin{equation}
\hat{W}\cos\theta+\left(\frac{\hat{U}+\hat{V}}{2}\right)\sin\theta
-\cos\theta\sin\theta=0,
\label{eq:kinconin01}
\end{equation}
\begin{equation}
-W\sin\theta+U\cos\theta
=
-\hat{W}\sin\theta+\hat{U}\cos\theta,
\label{eq:contvelbc01}
\end{equation}
\begin{equation}
-W\sin\theta+V\cos\theta
=
-\hat{W}\sin\theta+\hat{V}\cos\theta,
\label{eq:contvelbc02}
\end{equation}
\begin{equation}
\left(
\frac{\partial}{\partial r}-1
\right)\left(
-W\sin\theta+U\cos\theta
\right)=\lambda
\left(
\frac{\partial}{\partial r}-1
\right)\left(
-\hat{W}\sin\theta+\hat{U}\cos\theta
\right),
\label{eq:stressbc01}
\end{equation}
\begin{equation}
\left(
\frac{\partial}{\partial r}-1
\right)\left(
-W\sin\theta+V\cos\theta
\right)=\lambda
\left(
\frac{\partial}{\partial r}-1
\right)\left(
-\hat{W}\sin\theta+\hat{V}\cos\theta
\right).
\label{eq:stressbc02}
\end{equation}
As for the pressure field, we define $P$ and $\hat{P}$ as
\be
P=-\frac{p}{\cos\phi},\ \ \ 
\hat{P}=-\frac{\hat{p}}{\cos\phi}.
\ee
Since in the creeping flow approximation the pressure is harmonic, we can give a non divergent  solution  in a series of the associated Legendre functions $P_n^m$:
\begin{equation}
\begin{split}
P=&\sum_{n=1}^\infty\frac{A_nP_{n}^1(\mu)}{r^{n+1}},\ \ \ 
\hat{P}=\sum_{n=1}^\infty\hat{A}_n r^n P_{n}^1(\mu), 
\label{eq:pressure01}
\end{split}
\end{equation}
where $\mu=\cos\theta$ and $A_{n}$, $\hat{A}_{n}$ are unknown constants to be determined with the boundary conditions. Here we employ the definition of the associated Legendre functions defined in terms of the polynomials of zeroth order as $P_n^m(x)$$=(1-x^2)^{m/2}$ ${\rm d}^mP_n(x)/{\rm d}x^m$ (e.g. $P_1^1(\cos\theta)=\sin\theta$). From the Stokes equations (\ref{STOKESQ}), we can determine the general solutions in the expansion form
\begin{align}\nonumber
W=&
\sum_{n=1}^\infty\frac{nA_n}{2(2n+1)r^n}P_{n+1}^1(\mu)
+
\sum_{n=2}^\infty\frac{C_n}{r^n}P_{n-1}^1(\mu),
\nonumber\\
U=&
\sum_{n=1}^\infty\frac{A_n}{2(2n+1)r^n}P_{n+1}^2(\mu)
+
\sum_{n=3}^\infty\frac{E_n}{r^n}P_{n-1}^2(\mu),
\nonumber\\
V=&
-\sum_{n=1}^\infty\frac{n(n+1)A_n}{2(2n+1)r^n}P_{n+1}(\mu)
+
\sum_{n=1}^\infty\frac{G_n}{r^n}P_{n-1}(\mu),
\nonumber\\
\hat{W}=&
\sum_{n=2}^\infty \frac{(n+1)\hat{A}_nr^{n+1}}{2(2n+1)}P_{n-1}^1(\mu)
+
\sum_{n=0}^\infty\hat{C}_n r^{n+1}P_{n+1}^1(\mu),
\nonumber\\
\hat{U}=&
-\sum_{n=3}^\infty\frac{\hat{A}_nr^{n+1}}{2(2n+1)}P_{n-1}^2(\mu)
+
\sum_{n=1}^\infty\hat{E}_n r^{n+1}P_{n+1}^2(\mu),
\nonumber\\
\hat{V}=&
\sum_{n=1}^\infty\frac{n(n+1)\hat{A}_nr^{n+1}}{2(2n+1)}P_{n-1}(\mu)
+
\sum_{n=0}^\infty\hat{G}_n r^{n+1}P_{n+1}(\mu),\nonumber
\end{align}
with $C_{n}$, $E_{n}$, $G_{n}$ and $\hat{C}_{n}$, $\hat{E}_{n}$, $\hat{G}_{n}$ additional constants. Upon using the no-slip boundary condition on the plane wall (\ref{noslip})-(\ref{noslipb}) we get:
 \begin{align}
C_{2n}=&\frac{2n+1}{2(4n+1)}A_{2n},\ \ \ 
E_{2n-1}=\frac{(2n+1)A_{2n-1}}{4(n-1)(4n-1)},\nonumber\\
G_{2n-1}=&-\frac{(2n-1)^2A_{2n-1}}{2(4n-1)},\ \ \ 
\hat{C}_{2n}=\frac{n\hat{A}_{2n}}{4n+1},\nonumber\\
\hat{E}_{2n-1}=&-\frac{(n-1)\hat{A}_{2n-1}}{(2n+1)(4n-1)},\ \ \ 
\hat{G}_{2n-1}=\frac{2n^2\hat{A}_{2n-1}}{4n-1}, \nonumber
\end{align}
and, imposing the continuity equation (\ref{CONTQ}) together with (\ref{noslip})-(\ref{noslipb}), we find:
$$C_{2n-1}=\frac{(2n-1)(2n+1)}{4(4n-1)(n-1)}A_{2n-1},$$
$$E_{2n}=\frac{1}{2(n-1)(2n-1)}\left\{\frac{(2n-3)(2n+1)}{4n+1}A_{2n}+G_{2n}\right\},$$
$$\hat{C}_{2n-1}=\frac{2n(n-1)}{(2n+1)(4n-1)}\hat{A}_{2n-1},$$
$$\hat{E}_{2n}=\frac{1}{2(n+1)(2n+3)}\left\{-\frac{4n(n+2)}{4n+1}\hat{A}_{2n}+\hat{G}_{2n}\right\},$$
while, for the low-order coefficients, we obtain 
$$
A_1=G_1=\hat{E}_1=\hat{C}_0=\hat{C}_1=0,
$$
and
$$
\hat{A}_1 \neq 0,\ \ \ 
\hat{G}_0\neq 0.
$$
Using all these recursion relations we can rewrite the general solutions for the outer (${\bf v}=(v_{r},v_{\theta},v_{\phi})$) and inner velocities ($\hat{\bf v}=(\hat{v}_{r},\hat{v}_{\theta},\hat{v}_{\phi})$)  representing the solution of the equation set (\ref{eq:ge_cont02})-(\ref{eq:bc01}):
\be\label{eq:end1}
v_r=\left\{r\cos\theta\sin\theta
-W\cos\theta-\left(\frac{U+V}{2}\right)\sin\theta
\right\}\cos\phi,
\ee
\be\label{eq:end2}
v_\theta=\left\{r\cos^2\theta
+W\sin\theta-\left(\frac{U+V}{2}\right)\cos\theta
\right\}\cos\phi,
\ee
\be\label{eq:end3}
v_\phi=\left(-r\cos\theta
-\frac{U}{2}+\frac{V}{2}
\right)\sin\phi,
\ee
\be\label{eq:end4}
\hat{v}_r=\left\{r\cos\theta\sin\theta
-\hat{W}\cos\theta-\left(\frac{\hat{U}+\hat{V}}{2}\right)\sin\theta
\right\}\cos\phi,
\ee
\be\label{eq:end5}
\hat{v}_\theta=\left\{r\cos^2\theta
+\hat{W}\sin\theta-\left(\frac{\hat{U}+\hat{V}}{2}\right)\cos\theta
\right\}\cos\phi,
\ee
\be\label{eq:end6}
\hat{v}_\phi=\left(-r\cos\theta
-\frac{\hat{U}}{2}+\frac{\hat{V}}{2}
\right)\sin\phi
\ee
where
\begin{equation}
\begin{split}
W=&
\frac{3A_2P_1^1(\mu)}{10r^2}+
\frac{15A_3P_2^1(\mu)}{28r^3}\\&
+\sum_{n=2}^{\infty}
\left[
\left(
\frac{(n-1)A_{2n-2}}{(4n-3)r^{2n-2}}
+
\frac{(2n+1)A_{2n}}{2(4n+1)r^{2n}}\right)P_{2n-1}^1(\mu)
\right.\\&+\left.\left(
\frac{(2n-1)A_{2n-1}}{2(4n-1)r^{2n-1}}
+
\frac{(2n+1)(2n+3)A_{2n+1}}{4n(4n+3)r^{2n+1}}\right)P_{2n}^1(\mu)
\right],
\label{eq:gen_out_w01}
\end{split}
\end{equation}
\begin{equation}
\begin{split}
U=&
\frac{5A_3P_2^2(\mu)}{28r^3}+
\sum_{n=2}^{\infty}
\left[
\left(
\frac{A_{2n-2}}{2(4n-3)r^{2n-2}}
\right.\right.\\&\left.\left.
+
\frac{(2n-3)(2n+1)A_{2n}}{2(n-1)(2n-1)(4n+1)r^{2n}}
+
\frac{G_{2n}}{2(n-1)(2n-1)r^{2n}}\right)P_{2n-1}^2(\mu)
\right.\\&\left.
+
\left(
\frac{A_{2n-1}}{2(4n-1)r^{2n-1}}
+
\frac{(2n+3)A_{2n+1}}{4n(4n+3)r^{2n+1}}\right)P_{2n}^2(\mu)
\right],
\label{eq:gen_out_u01}
\end{split}
\end{equation}
\begin{equation}
\begin{split}
V=&
\frac{G_2P_1(\mu)}{r^2}
-
\frac{9A_3P_2(\mu)}{14r^3}\\&
+\sum_{n=2}^{\infty}
\left[
\left(
-\frac{(n-1)(2n-1)A_{2n-2}}{(4n-3)r^{2n-2}}
+\frac{G_{2n}}{r^{2n}}\right)P_{2n-1}(\mu)
\right.\\&\left.
+\left(
-\frac{n(2n-1)A_{2n-1}}{(4n-1)r^{2n-1}}
-\frac{(2n+1)^2A_{2n+1}}{2(4n+3)r^{2n+1}}\right)P_{2n}(\mu)
\right],
\label{eq:gen_out_v01}
\end{split}
\end{equation}
\begin{equation}
\begin{split}
\hat{W}=&
\frac{3\hat{A}_{2}r^3P_1^1(\mu)}{10}
\\&
+\sum_{n=2}^\infty
\left(
\frac{(2n+1)\hat{A}_{2n}r^{2n+1}}{2(4n+1)}
+
\frac{(n-1)\hat{A}_{2n-2}r^{2n-1}}{4n-3}
\right)P_{2n-1}^1(\mu)
\\&
+\sum_{n=1}^\infty
\left(
\frac{(n+1)\hat{A}_{2n+1}r^{2n+2}}{4n+3}
+
\frac{2n(n-1)\hat{A}_{2n-1}r^{2n}}{(2n+1)(4n-1)}
\right)P_{2n}^1(\mu),
\label{eq:gen_in_w01}
\end{split}
\end{equation}
\begin{equation}
\begin{split}
\hat{U}=&
\sum_{n=2}^\infty
\left(
-\frac{\hat{A}_{2n}r^{2n+1}}{2(4n+1)}
-\frac{2(n-1)(n+1)\hat{A}_{2n-2}r^{2n-1}}{n(2n+1)(4n-3)}
\right.\\&\left.
+\frac{\hat{G}_{2n-2}r^{2n-1}}{2n(2n+1)}
\right)
P_{2n-1}^2(\mu)
\\&+
\sum_{n=1}^\infty
\left(
-\frac{\hat{A}_{2n+1}r^{2n+2}}{2(4n+3)}
-\frac{(n-1)\hat{A}_{2n-1} r^{2n}}{(2n+1)(4n-1)}
\right)P_{2n}^2(\mu),
\label{eq:gen_in_u01}
\end{split}
\end{equation}
\begin{equation}
\begin{split}
\hat{V}=&
\frac{\hat{A}_1r^2P_0(\mu)}{3}+
\sum_{n=1}^\infty
\left[
\left(
\frac{n(2n+1)\hat{A}_{2n}r^{2n+1}}{4n+1}
+
\hat{G}_{2n-2} r^{2n-1}
\right)P_{2n-1}(\mu)
\right.\\&+\left.
\left(
\frac{(n+1)(2n+1)\hat{A}_{2n+1}r^{2n+2}}{4n+3}
+
\frac{2n^2\hat{A}_{2n-1}r^{2n}}{4n-1}
\right)P_{2n}(\mu)
\right].
\label{eq:gen_in_v01}
\end{split}
\end{equation}
In the previous expressions the six unknown sets of coefficients $A_{2n+1}$, $A_{2n}$, $G_{2n}$, $\hat{A}_{2n-1}$, $\hat{A}_{2n}$ and $\hat{G}_{2n-2}$ with $n \ge 1$ have now to be determined by the imposition of the six conditions (\ref{eq:kinconou01})-(\ref{eq:stressbc02}) on the hemispherical bump.

\subsection{Determination of the coefficients}

To determine the coefficients  $A_{2n+1}$, $A_{2n}$, $G_{2n}$, $\hat{A}_{2n-1}$, $\hat{A}_{2n}$ and $\hat{G}_{2n-2}$ for $n\geq 1$, we need recurrence formulae that can be deduced from the boundary conditions at $r=1$  given by 
(\ref{eq:kinconou01})-(\ref{eq:stressbc02}).  These recurrence formulae are given in the appendix  and, although they are written as infinite series, 
we numerically solve the problem by truncating the system of equations determining the coefficients. For each of the six coefficients we consider modes up to $n=100$ and to have a good indication of the accuracy of this procedure, the numerically estimated coefficients are substituted into (\ref{eq:kinconou01})-(\ref{eq:stressbc02}) until the boundary conditions are satisfied to at least five decimal places. The coefficients $A_{2n+1}$, $A_{2n}$, $G_{2n}$, $\hat{A}_{2n-1}$,
$\hat{A}_{2n}$ and $\hat{G}_{2n-2}$ up to $n=15$
are listed in tables \ref{tab:coef_chi0}-\ref{tab:coef_chi1e100} for $\lambda=0$, $0.1$, $1$, $10$ and $10^{100}$ and  to obtain the full solutions, they must be used into equations (\ref{eq:end1})-(\ref{eq:gen_in_v01}) \footnote{For the numerical coefficients corresponding to other values of $\lambda$ the interested reader can contact the authors}. 

\begin{figure}
\begin{center}
{\includegraphics[scale=0.5,angle=0]{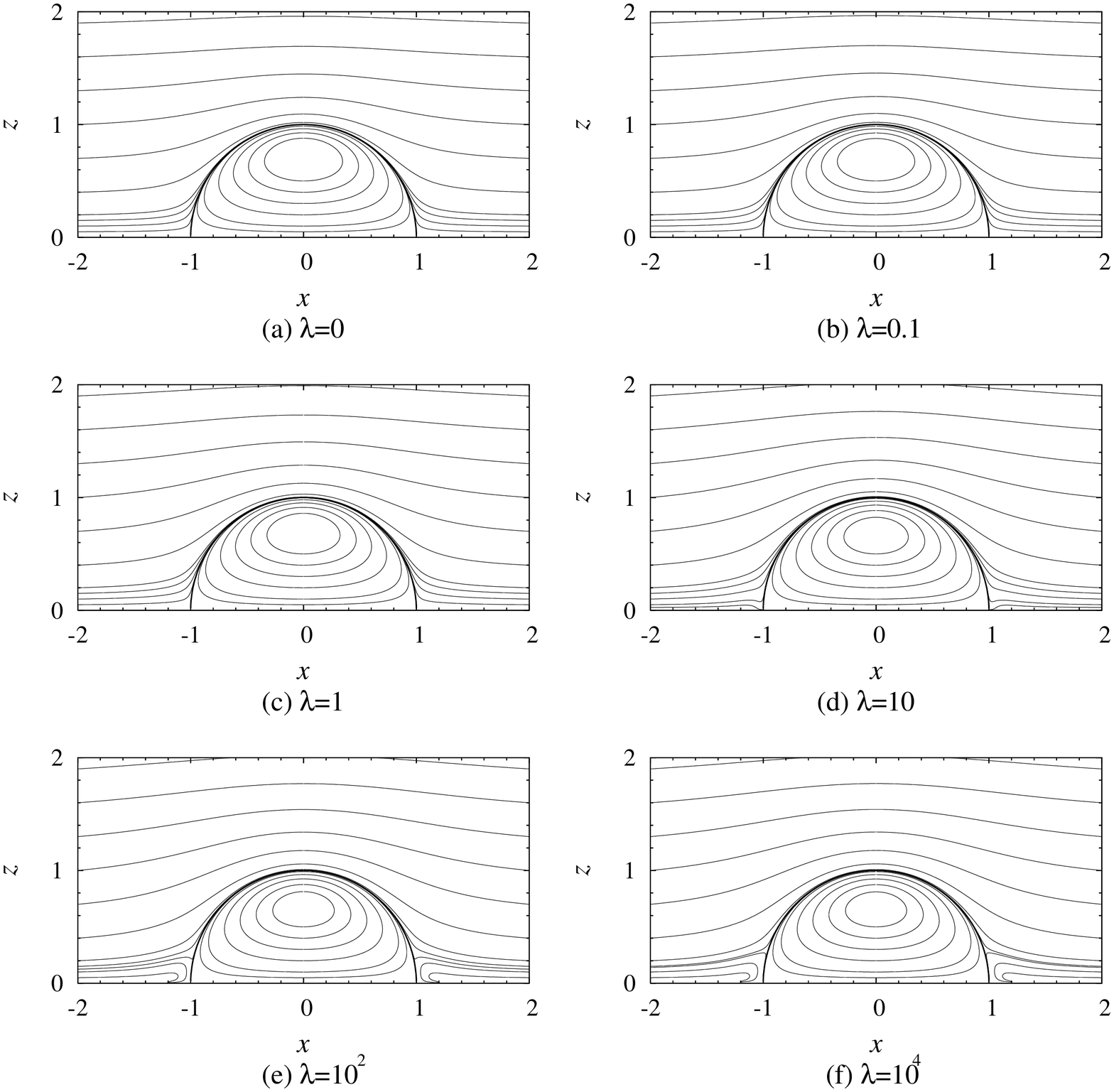}}
\caption{Streamlines patterns in the in $xz$ mid-plane (for $y=0$).}
\label{fig:2}
\end{center}
\end{figure}

In figure \ref{fig:2} we show the kinematical structure of the flow and the development of stagnation points and flow reversal. Let us notice that for high viscosity ratio the streamlines plot is consistent with what obtained by Pozrikidis \cite{Pozrikidis}.\\
It is now useful to check the validity for two extreme limits $\lambda\rightarrow 0$ and $\lambda \rightarrow \infty$.
For the free-slip bump, $\lambda=0$, as also confirmed by our numerical solution shown in Table \ref{tab:coef_chi0},   the solution is analytical and can be expressed as
\be\nonumber
A_n=\frac{2}{3}\delta_{n2},\ \ \ 
G_n=\frac{2}{5}\delta_{n2}.
\ee
Using this solution, we determine $U$, $V$, $W$ and $P$ 
\begin{equation}\nonumber
\begin{split}
W=&\frac{\sqrt{1-\mu^2}\mu^2}{r^2},\ \ \ 
U=\frac{(1-\mu^2)\mu}{r^2},\ \ \   
V=\frac{(1-\mu^2)\mu}{r^2},\\
P=&\frac{2\sqrt{1-\mu^2}\mu}{r^3}.
\end{split}
\end{equation}
The velocity field around the hemisphere is then given by
\begin{equation}\nonumber
\begin{split}
{\bf v}=&
{\bm e}_r \left(r-\frac{1}{r^2}\right)\cos\theta\sin\theta\cos\phi
+
{\bm e}_\theta r\cos^2\theta\cos\phi
-
{\bm e}_\phi r\cos\theta\sin\phi
\label{eq:tildq01}
\end{split}
\end{equation}
that can also be rewritten together with the pressure in Cartesian coordinates
\begin{equation}
\begin{split}
{\bf v}=&
{\bm e}_x\left(z-\frac{x^2z}{r^5}\right)
-{\bm e}_y\frac{xyz}{r^5}
-{\bm e}_z\frac{xz^2}{r^5},\\
p=&-\frac{2xz}{r^5}.
\label{eq:tildq02}
\end{split}
\end{equation}
It should be noted that the present solution (\ref{eq:tildq02})
is same as the flow around a free-slip sphere in an infinite fluid. 
It is because  the presence of the free-slip sphere does not alter
the $\theta$ and $\phi$ components of the velocity vector 
from the uniform shear flow and then the modulated velocity 
becomes zero at $z=0$.\\
In the case of the largest viscosity ratio $\lambda=10^{100}$, 
the coefficients $A_{2n+1}$, $A_{2n}$ and $G_{2n}$ 
listed in Table \ref{tab:coef_chi1e100} show quantitative agreement with those 
for the no-slip bump shown in the paper by Price \cite{Price}. 
The inner solution is analytical and given by $\hat{A}_n=0$ and $\hat{G}_n=2\delta_{0n}$ that imply a zero velocity inside the bump.\\

\subsection{Torque and Force}

\begin{figure}
\begin{center}
\includegraphics[width=.5\textwidth,angle=270]{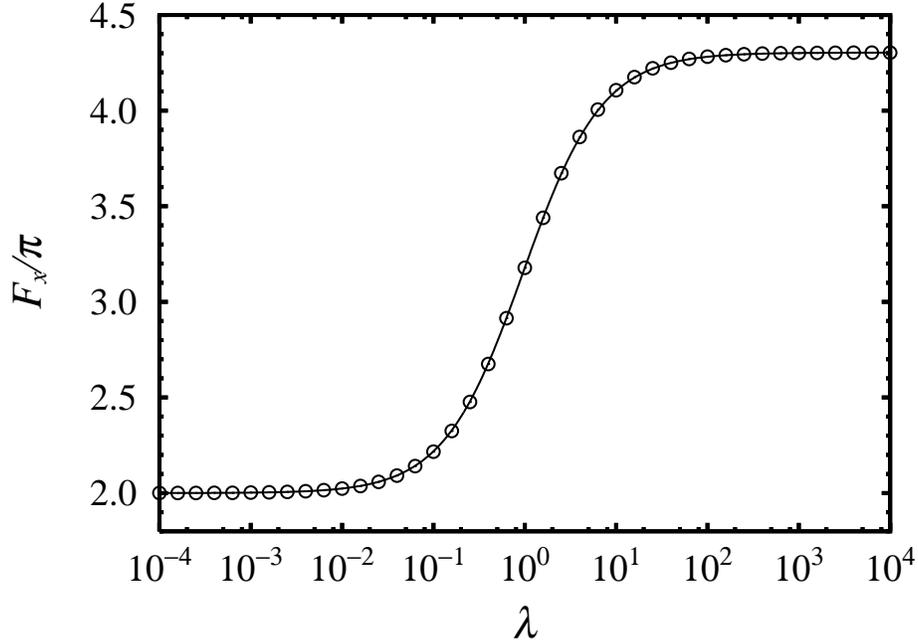}
\caption{
The force acting on the hemisphere versus 
the viscosity ratio.  The symbol $\circ$ corresponds to the evaluation based on (\ref{eq:force03}). The solid line corresponds to the approximation (\ref{eq:approx_fx}).}
\label{fig:3}
\end{center}
\end{figure}

\begin{figure}
\begin{center}

\includegraphics[width=.5\textwidth,angle=270]{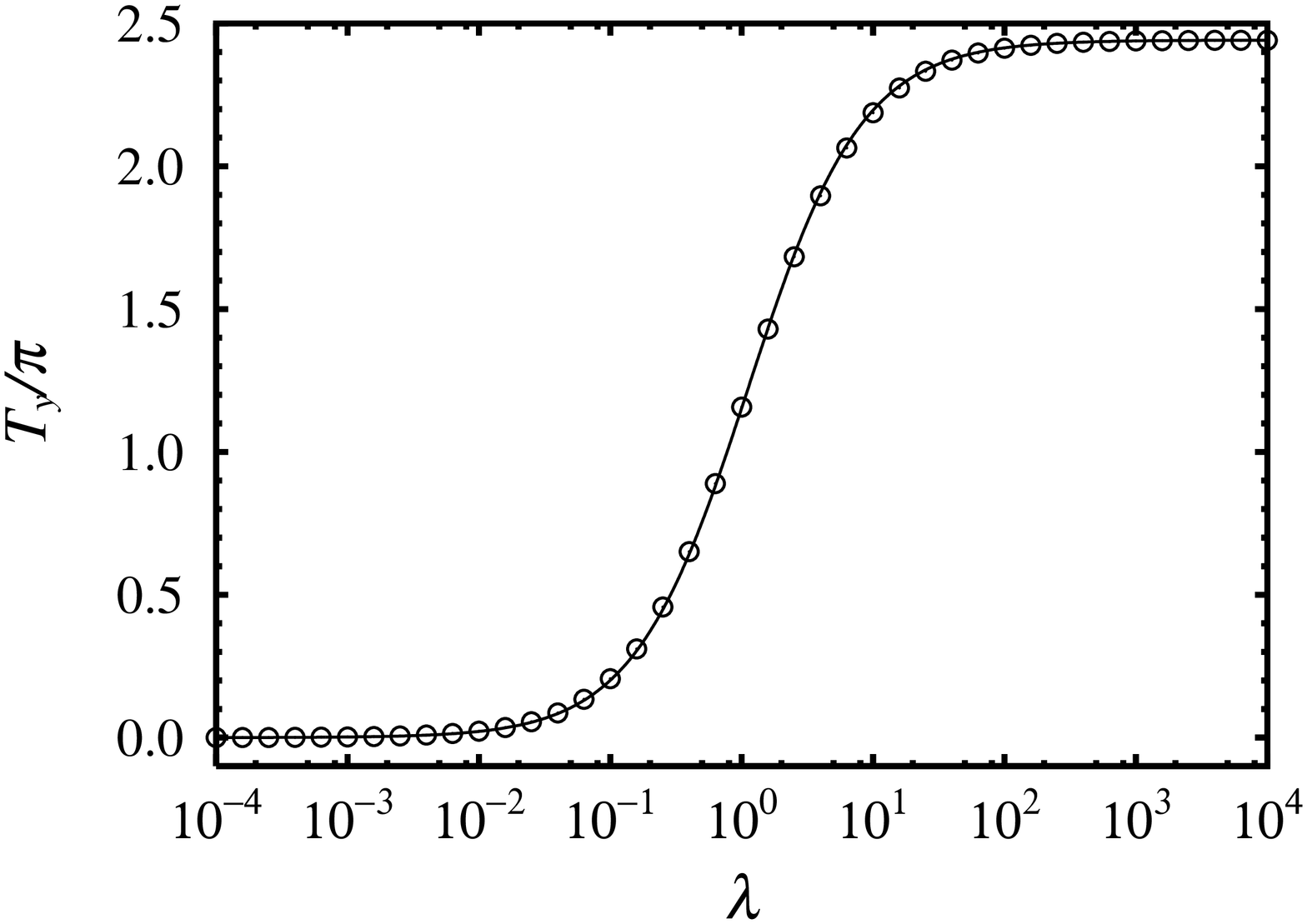}
\caption{ The torque acting on the hemisphere versus 
the viscosity ratio.  The symbol $\circ$ corresponds to the evaluation based on (\ref{eq:torque03}). The solid line corresponds to the approximation (\ref{eq:approx_ty}).
}
\label{fig:4}
\end{center}
\end{figure}

We now evaluate the force and the torque acting on the hemisphere for various viscosity ratios. The general strategy will be to write them as a functions of the constants $A_{2n},S_{2n+1},G_{2n}$ and then use the coefficients obtained from the numerical solution in the previous section in order to evaluate them. To do this, we write the force ${\bm F}$ and the torque ${\bm T}$  in the most  general form
\begin{align}
{\bm F}=&
\int_0^{2\pi}\!\!\!{\rm d}\phi
\int_0^{\pi/2}\!\!\!{\rm d}\theta\ 
\sin\theta\ ({\bm \sigma}\cdot{\bm e}_r)_{r=1},\nonumber
\\
{\bm T}=&
\int_0^{2\pi}\!\!\!{\rm d}\phi
\int_0^{\pi/2}\!\!\!{\rm d}\theta\ 
\sin\theta\ [r{\bm e}_r\times({\bm \sigma}\cdot{\bm e}_r)]_{r=1}.\nonumber
\end{align}
Let us notice that the symmetry of the system with respect to $\phi$ is such that only the $x$ component of the planar force vector $(F_x,F_{y})$ and 
$y$ component of the torque vector $T_y$ are non-zero. 
To estimate these components, considering the kinetic condition in (\ref{eq:bc02})
and using $U$, $V$ and $W$ given by the general solution we can exactly express the torque and force as a function of $A_{2n}$,$A_{2n+1}$ and $G_{2n}$:
\begin{equation}\label{eq:force03}
\begin{split}
F_x=&\pi\left[
\frac{1}{2}
+\frac{4}{5}A_3
-\sum_{n=1}^\infty
\frac{(-1)^{n}(2n+1)!!}{(2n-3)(2n-1)(2n-2)!!}
\right.\\&\left.\times
\left(
\frac{(4n^3-9n-2)A_{2n}}
{2(n+1)(4n+1)}
+
\frac{(2n^2-3n-1)G_{2n}}
{n(2n+1)}
\right)
\right],
\end{split}
\end{equation}
\begin{equation}\label{eq:torque03}
\begin{split}
T_y=&
\pi\left[
-\frac{3}{5}A_2+G_2
+\sum_{n=1}^\infty
\frac{(-1)^{n}(2n+1)!!}{(2n-1)(2n)!!}
A_{2n+1}\right],
\end{split}
\end{equation}
with
$$(2n+1)!!=(2n+1)\times(2n-1)\times...\times 1,$$
$$(2n)!!=(2n)\times(2n-2)\times...\times 2,\hspace{.2in} 0!!=1.$$
In the limit of highly viscous droplet ($\lambda \rightarrow \infty$), from the numerical evaluation of the coefficients for $\lambda=10^{100}$  (see  Table \ref{tab:coef_chi1e100}) we determine the force and torque as
\be\nonumber
F_x=4.30322\pi, \ \ \ T_y=2.44132\pi,
\ee
and both results are consistent with the evaluations $F_x=4.30\pi$,$T_y=2.44\pi$ for the rigid no-slip bump given in the paper by Price \cite{Price}. \\
The profiles of $F_x$ and $T_y$ as a function of $\lambda$ are respectively shown in figure \ref{fig:3} and figure \ref{fig:4}. 
As is clearly shown both the torque and the force approach a finite value for very small ($\lambda\rightarrow 0$) and large ($\lambda\rightarrow \infty$) viscosity ratio. Considering these asymptotic behaviors, we can estimate
\begin{align}
\frac{F_x}{\pi}\approx&\frac{2+4.51003\lambda}{1+1.04806\lambda},
\label{eq:approx_fx}
\\
\frac{T_y}{\pi}\approx&\frac{2.18808\lambda}{1+0.896271\lambda},
\label{eq:approx_ty}
\end{align}
in which the numerical coefficients have been determined from the method of least squares.

\subsection{Effects of Capillary number on deformation}

In this section we study the local difference of the normal stress between the inner and outer regions of the bump. If we assume a finite surface tension  $\gamma$ at the droplet surface, the difference between the the two normal stresses must equal the surface tension times the local curvature of the hemispherical bump. In other words, for fixed surface tension, this gives the possibility to estimate the deformations on the droplet surface consistently with the Laplace law \cite{Chan,Magnaudet}. Resting on the assumption that these deformations are small, we can introduce a Capillary number as 
\be\nonumber
Ca=\frac{\eta S R_{0}}{\gamma},
\ee
where $R_{0}$, $S$ and $\gamma$ are dimensional length, shear and surface tension scales. Next we write a dimensional stress tensor as
\be\nonumber
{{\bm \sigma}}^{(dim)}={{\bm \sigma}} S \eta-P_{s},
\ee
\be\nonumber
{\hat{\bm \sigma}}^{(dim)}= {\hat{\bm \sigma}} S {\eta} \lambda-\hat{P}_{s},
\ee
where the contributions ${{\bm \sigma}}$ and ${\hat{\bm \sigma}}$ are purely dimensionless, i.e. they correspond to a unitary shear $S$ and unitary viscosity $\eta$ described in the previous sections. The terms $P_{s}$ and $\hat{P}_{s}$ represent constant dimensional background pressures, i.e. the pressure inside/outside the bump in absence of flow. Assuming $Ca\ll 1$, if we indicate with $r=R$ the position of the interface, we should write: 
\be
R(\theta,\phi)=R_{0}(1+Ca R^{(1)}(\theta,\phi)+{\cal O}(Ca^2)),
\label{eq:deform_r01}
\ee
where $R^{(1)}$ accounts for the displacement from the pure hemisphere. The Laplace law on the interface is written in a general form
\be\nonumber
\kappa^{(dim)} \gamma= {\sigma}^{(dim)}_{rr}-\hat{{\sigma}}^{(dim)}_{rr},
\ee
where $\kappa^{(dim)}=\frac{\kappa}{R_{0}}$ denotes the dimensional curvature, with $\kappa$ its dimensionless counterpart. If $Ca\ll 1$, the displacement is  small enough and the curvature on the bump interface is given by 
\begin{equation}\nonumber
\begin{split}
\kappa=&
R_{0}\left\{\nabla\cdot \left(\frac{\nabla(r-R)}{|\nabla(r-R)|}\right)\right\}_{r=R}\approx
R_{0} \left\{
\nabla^2 (r-R)
\right\}_{r=R}
\\=&
\frac{2}{1+Ca R^{(1)}+...}
-
\left[
\frac{1}{\sin\theta}
\frac{\partial}{\partial\theta}
\left(\sin\theta
\frac{\partial}{\partial\theta}\right)
+\frac{1}{\sin^2\theta}
\frac{\partial^2}{\partial \phi^2}
\right]Ca R^{(1)}+...
\\=&
2Ca^{0}
-
Ca^{1}
\left[
2+
\frac{1}{\sin\theta}
\frac{\partial}{\partial\theta}
\left(\sin\theta
\frac{\partial}{\partial\theta}\right)
+\frac{1}{\sin^2\theta}
\frac{\partial^2}{\partial \phi^2}
\right]R^{(1)}+{\cal O}(Ca^2),
\end{split}
\end{equation}
hence $\frac{\kappa}{Ca}= \left( {\sigma}_{rr}-\lambda \hat{{\sigma}}_{rr} \right)_{r=1}+(\hat{P}_{s}-P_{s})/(S \eta)$ and
\begin{equation}
\begin{split}
\label{Laplaceext}
&\frac{2}{Ca}-\left[
2+
\frac{1}{\sin\theta}
\frac{\partial}{\partial\theta}
\left(\sin\theta
\frac{\partial}{\partial\theta}\right)
+\frac{1}{\sin^2\theta}
\frac{\partial^2}{\partial \phi^2}
\right]R^{(1)}+{\cal O}(Ca^1)
\\
=&\left( {\sigma}_{rr}-\lambda \hat{{\sigma}}_{rr} \right)_{r=1}+(\hat{P}_{s}-P_{s})/(S \eta).
\end{split}
\end{equation}
In the absence of the shear flow, the pressure is constant inside and outside the bump and equal to $\hat{P}_{s}$ and $P_{s}$ respectively. The gap between the two pressures  is determined by $(\hat{P}_{s}-P_{s})/(S \eta)=2Ca^{-1}$ from the ${\cal O}(Ca^{-1})$ of (\ref{Laplaceext}).  This means that in  presence of a non zero velocity, in order to find consistency with the Laplace law, we should consider a finite displacement given by a non zero $R^{(1)}$ in (\ref{eq:deform_r01}). To determine this displacement, equation (\ref{Laplaceext}) is solved at ${\cal O}(Ca^0)$ :
\be
-\left[
2+
\frac{1}{\sin\theta}
\frac{\partial}{\partial\theta}
\left(\sin\theta
\frac{\partial}{\partial\theta}\right)
+\frac{1}{\sin^2\theta}
\frac{\partial^2}{\partial \phi^2}
\right]R^{(1)}
=\left(
{\sigma}_{rr}-\lambda\hat{{\sigma}}_{rr}
\right)_{r=1},
\label{eq:laplace_law01}
\ee
with the boundary condition 
\be\label{COND4}
R^{(1)}=0\ \ \ {\rm at}\ \ \ 
\theta=\frac{\pi}{2},0\le \phi \le 2 \pi.
\ee
This condition corresponds to fix the contact line while accounting for the shape deformation of the droplet. More generally, the contact line will be deformed and in principle one can solve for the contact line shape as part of the problem \cite{Dimi1,Dimi2}. However, the requirement (\ref{COND4}) is still useful, given the comparison it allows with other works  \cite{Li,Yon}.\\
If we write the normal stress jump in the expansion form with respect to the associated Legendre polynomials
\be\nonumber
\left({\sigma}_{rr}-\lambda\hat{{\sigma}}_{rr}
\right)_{r=1}=\sum_{n=1}^{\infty}Q_nP_n^1(\mu)\cos\phi,
\ee
the expansion coefficients $Q_n$ can be determined as:
\begin{equation}
\begin{split}
Q_{2n}=&
\frac{2\delta_{1n}}{3}(1-\lambda)
+
\frac{4n^2+6n-1}{4n-1}A_{2n}
\\&
+
\frac{2(n+1)^2(2n+3)(8n^2+6n-3)}{n(2n+1)(4n+3)(4n+5)}A_{2n+2}
+
\frac{2(n+1)}{n(2n+1)}G_{2n+2}
\\&+\lambda\left(
\frac{2(n-1)(2n-1)^2(4n^2+n-2)}{n(2n+1)(4n-3)(4n-1)}\hat{A}_{2n-2}
\right.\\ &\left.
\ \ \ \ \ \ 
+\frac{4n^2-2n-3}{4n+3}\hat{A}_{2n}
+\frac{2n-1}{n(2n+1)}\hat{G}_{2n-2}
\right),
\label{eq:q2n_01}
\end{split}
\end{equation}
\begin{equation}
\begin{split}
Q_{2n-1}=&
\frac{4n^2+2n-3}{4n-3}A_{2n-1}
+
\frac{(n+1)(2n+1)^2}{n(4n+1)}A_{2n+1}
\\&
+\lambda\left(
\frac{4(n-1)^2(2n-3)}{(2n-1)(4n-3)}\hat{A}_{2n-3}
+
\frac{4n^2-6n-1}{4n+1}\hat{A}_{2n-1}
\right).
\label{eq:q2nm1_01}
\end{split}
\end{equation}
We now solve (\ref{eq:laplace_law01}) and determine the displacement 
\begin{equation}
R^{(1)}=\left(
\frac{R_0^{(1)}\sqrt{1-\mu^2}}{1+\mu}
+
\sum_{n=2}^{\infty}R_n^{(1)}P_n^1(\mu)\right)
\cos\phi,
\label{eq:deform_r02}
\end{equation}
where 
\be\label{DEF1}
R_0^{(1)}=\sum_{k=2}^{\infty}\frac{(-1)^{k}(2k-1)!!}{2(k-1)(2k+1)(2k-2)!!}
Q_{2k-1},
\ee
\be\label{DEF2}
R_n^{(1)}=\frac{Q_n}{(n-1)(n+2)}\ \ \ 
{\rm for}\ \ \ n\geq 2.
\ee
Concerning the $\phi$ dependence of the normal stress jump, this is simply proportional to $\cos\phi$.  
\begin{figure}
\begin{center}
{\includegraphics[scale=0.5,angle=0]{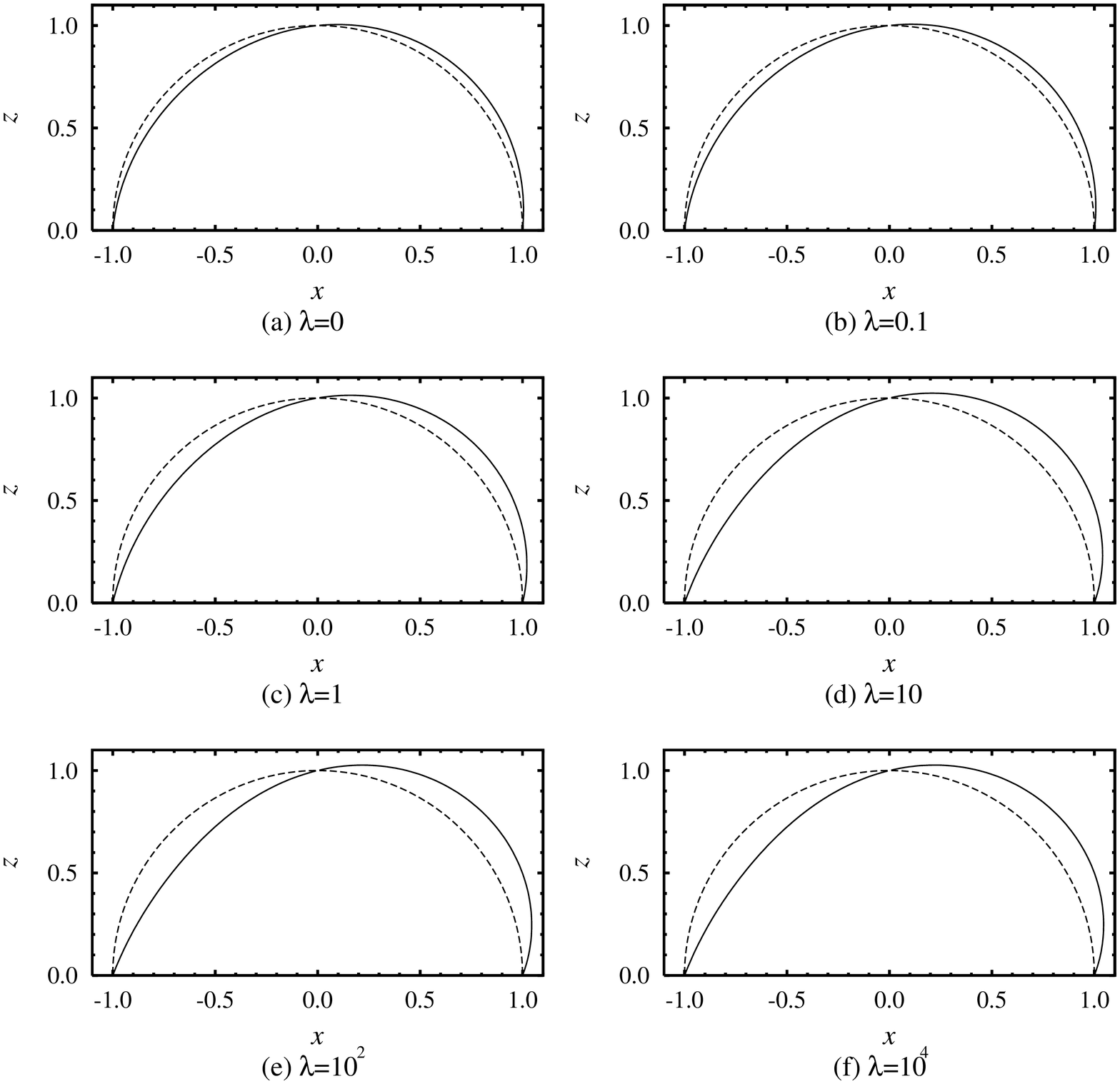}}
\caption{
Shape of the bump on the plane of $y=0$
for various viscosity ratios $\lambda$.
The dashed line shows the hemisphere with no deformation. 
The solid line shows the deformed interface 
for a Capillary number   $Ca=0.05$.
}
\label{fig:5}
\end{center}
\end{figure}
In figure \ref{fig:5} we show the deformation of the bump on the plane of $y=0$ with a Capillary number $Ca=0.05$. These deformations have been estimated from (\ref{eq:deform_r02}),(\ref{DEF1}) and (\ref{DEF2}) with the numerical coefficients in (\ref{eq:q2n_01}) and (\ref{eq:q2nm1_01}) determined with the procedure of the previous section. Moreover, to better quantify the role of deformations with respect to the hemispherical bump, advancing and receding  contact angles should be respectively defined by $\pi/2+\Delta\theta$ and $\pi/2-\Delta\theta$,  with $\Delta \theta$  proportional to $Ca$  if the  deformation is small enough (see figure \ref{fig:6}). 

\begin{figure}
\begin{center}
\includegraphics[width=.5\textwidth,angle=0]{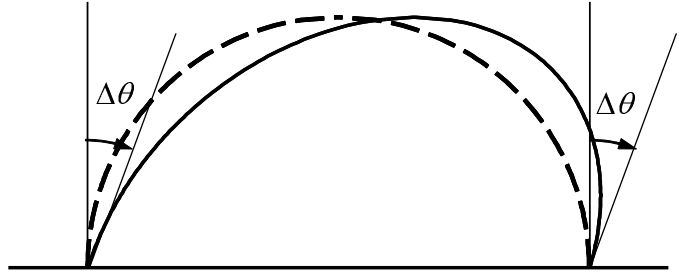}
\caption{Advancing and receding contact angles due to deformations are introduced in the problem in the small Capillary number regime.}
\label{fig:6}
\end{center}
\end{figure}

\begin{figure}
\begin{center}
\includegraphics[width=.5\textwidth,angle=270]{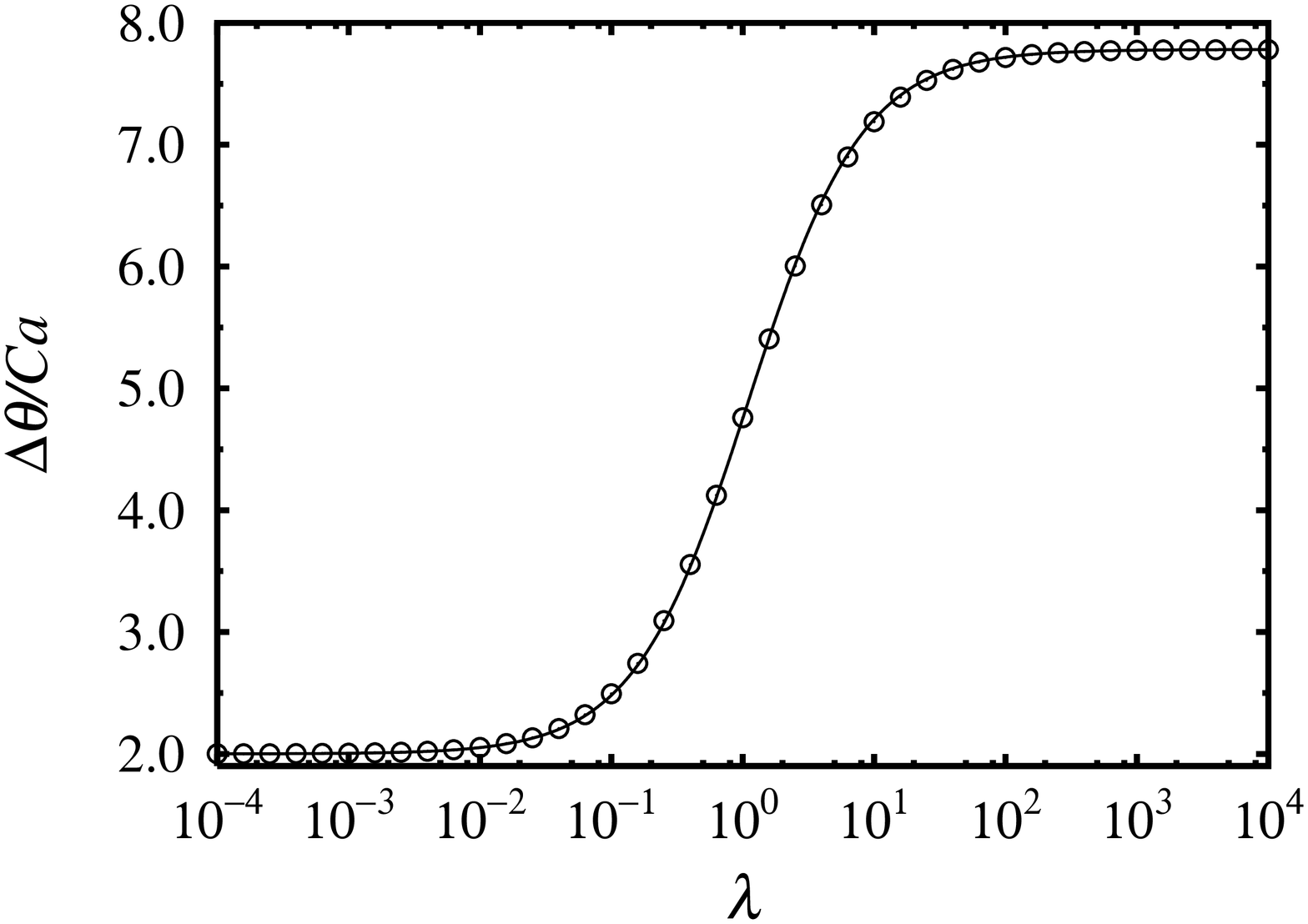}
\caption{Contact angle deformations as a function of the viscosity ratio in the limit of small Capillary numbers. The symbol $\circ$ corresponds to the solution based on (\ref{eq:laplace_law01}). The solid line corresponds to the approximation (\ref{Caapprox}).}
\label{fig:7}
\end{center}
\end{figure}

For various viscosity ratios $\lambda$, $\Delta \theta/Ca$ is then plotted in figure  \ref{fig:7}. Considering the asymptotic behaviors, this deformation can be approximated by
\be\label{Caapprox}
\Delta\theta
\approx
\frac{(2+7.07185\lambda)}{(1+0.908826\lambda)} Ca,
\ \ \ 
\ee
in which the numerical coefficients have been determined from the method of least squares. \\
For the case of hemispherical droplet whose viscosity is equal to the surrounding fluid ($\lambda=1$), it is instructive to compare our results with the ones of Li and Pozrikidis \cite{Li}. 

\begin{figure}
\begin{center}
\includegraphics[width=.5\textwidth,angle=270]{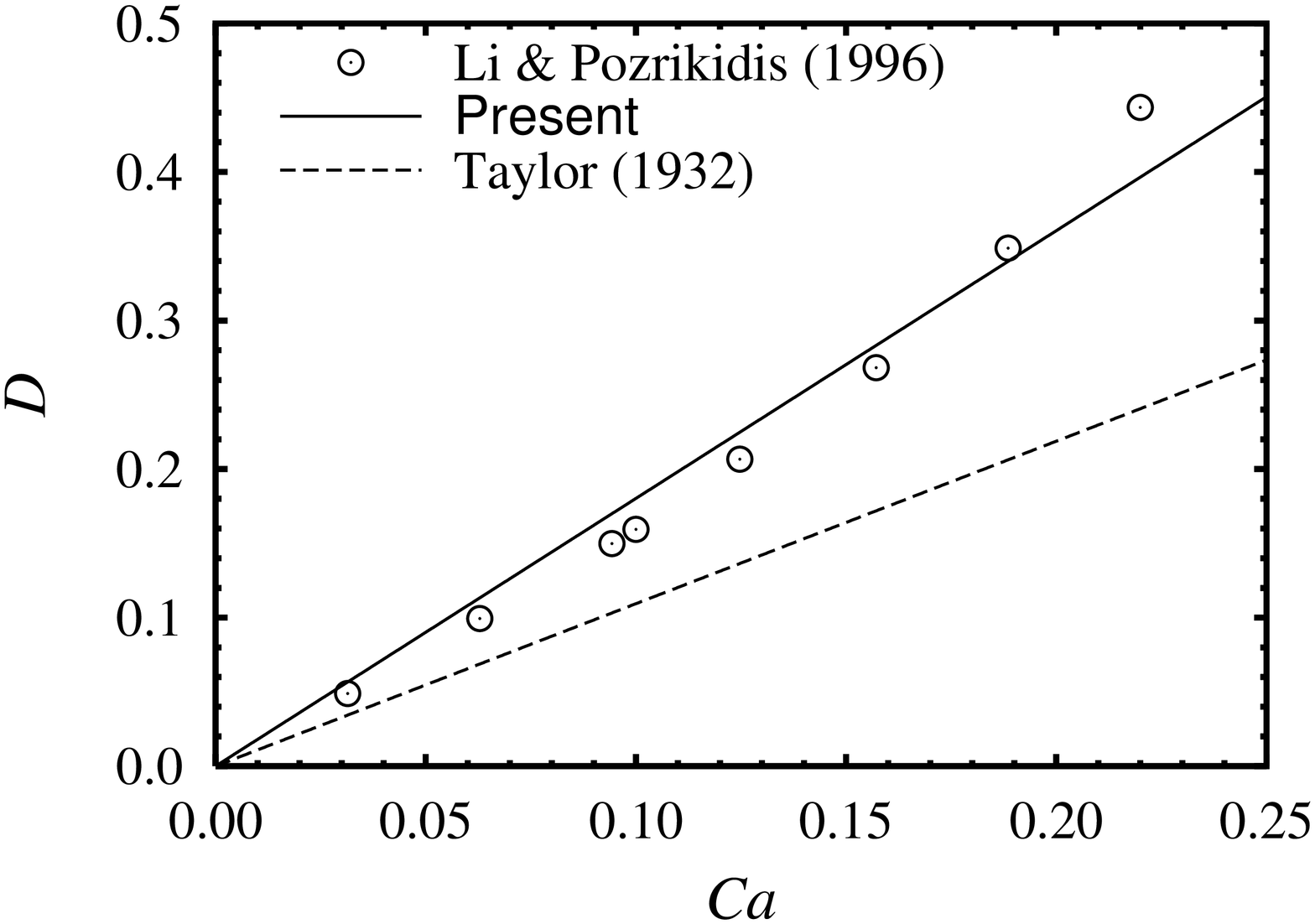}
\caption{Deformation parameter $D$ as a function of the Capillary number for a viscosity ratio $\lambda=1$. Our solution (straight line) is compared with the results  obtained with the boundary integral method ($\circ$) by Li and Pozrikidis (1996). The deformation parameter for the case of a droplet suspended in an infinite shear flow is also reported (dashed line).}
\label{fig:8}
\end{center}
\end{figure}

\begin{figure}
\begin{center}
\includegraphics[width=.5\textwidth,angle=270]{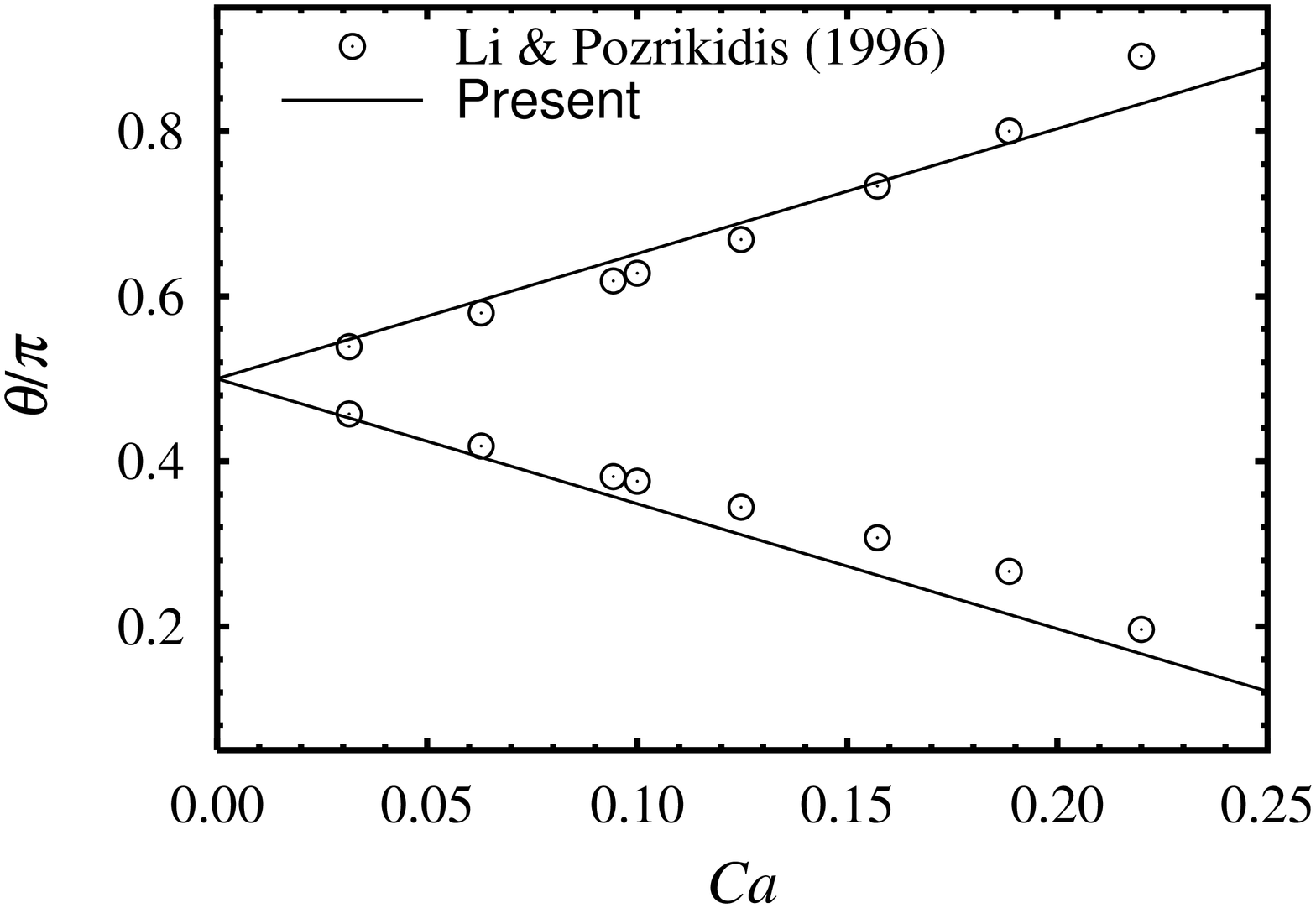}
\caption{Advancing and receding contact angles as a function of the Capillary number for a viscosity ratio $\lambda=1$. Our results (straight line) are compared with the numerical results obtained with the boundary integral method ($\circ$) by Li and Pozrikidis (1996).}
\label{fig:9}
\end{center}
\end{figure}
\begin{figure}
\begin{center}
\includegraphics[width=.5\textwidth,angle=270]{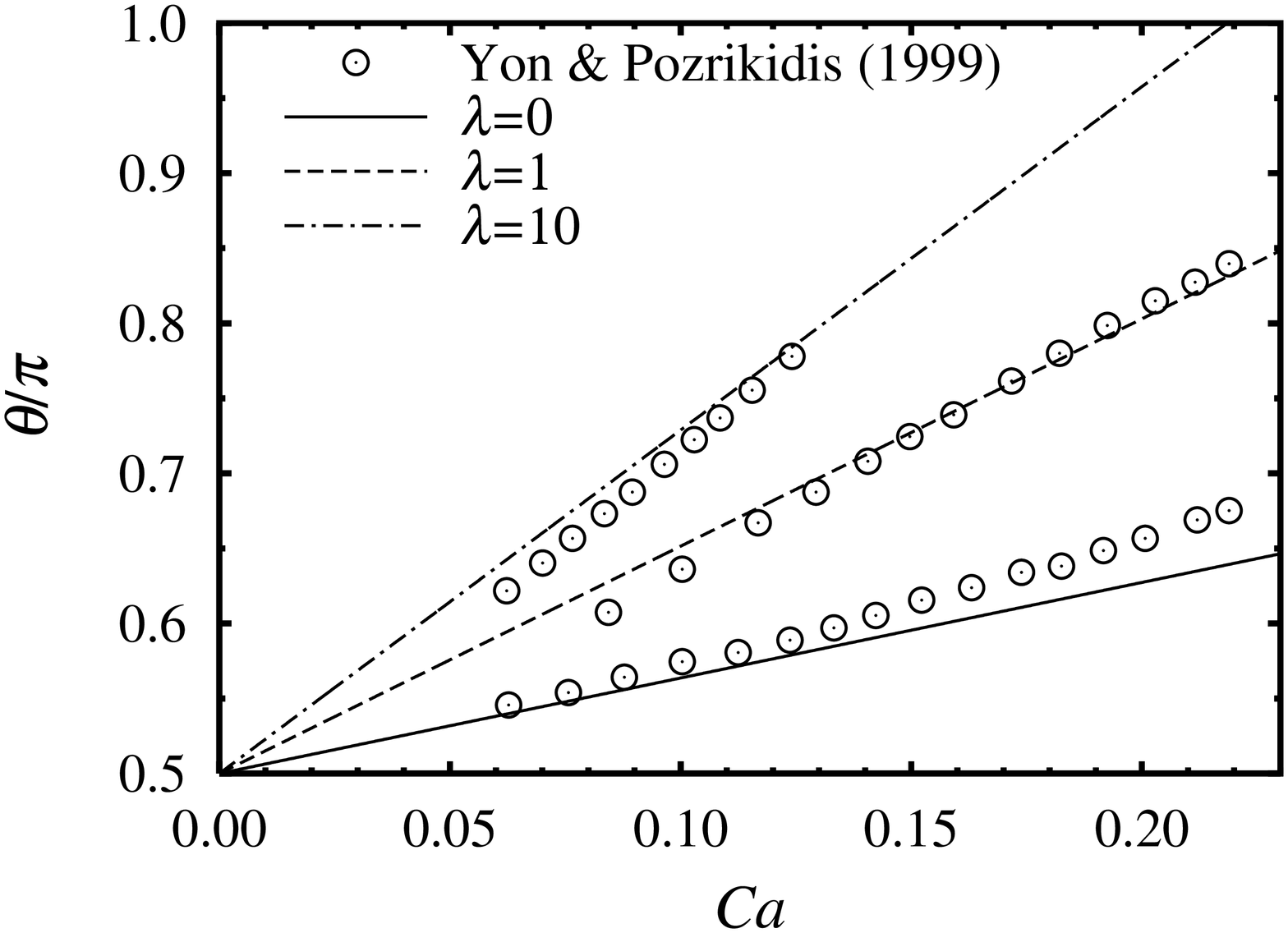}
\caption{Advancing contact angles as a function of the Capillary number for viscosity ratios $\lambda=0,1,10$. Our results are compared with the numerical results obtained with the boundary integral method ($\circ$) by Yon and Pozrikidis (1999).}
\label{fig:10}
\end{center}
\end{figure}

In figure \ref{fig:8} we plot the deformation parameter $D={A-B \over A+B}$ as a function of the Capillary number. $A$ and $B$ are, respectively, the maximum and minimum radial distances of the interface from the origin. Note that the corresponding slope for a  spherical drop suspended in an infinite simple shear flow is equal to $35/32$ \cite{Taylor} and suggests that the wall is promoting the deformation of the drop as it is also confirmed by the computation of Li and Pozrikidis \cite{Li} (see their figure 6 (e)) with the boundary integral method. To further compare our results with the ones of Li and Pozrikidis, in figure \ref{fig:9} we plot the advancing and receding contact angles as a function of the Capillary number and a good agreement is found.\\
The numerical investigations of Li and Pozrikidis \cite{Li} were also extended in another paper by Yon and Pozrikidis \cite{Yon} by examining the role played by the viscosity ratio. In particular in figure \ref{fig:10} we compare our result for  the advancing contact angle as a function of the Capillary number with the numerical results shown in figure 3 (g) of the paper by Yon and Pozrikidis and  quantitative agreement is found.

\section{Concluding remarks}

An exact solution for the problem of a linear shear flow overpassing a hemispherical droplet resting on a plane wall has been explicitly computed. The results here presented are the natural generalization of the ones proposed in an earlier paper by Price \cite{Price} who investigated the case of a hemispherical no-slip bump. Several expressions including  the torque and force acting on the hemisphere are exactly derived and computed as a function of the viscosity ratio. The small deformations on the hemispherical surface have also been quantified in the limit of small Capillary numbers and a comparison with available results has been presented. Extensions of the present study can be made into several directions. One extension concerns the generalization of the present results to non hemispherical droplets \cite{Pozrikidis,Yon}. In this case, one would apply the same procedure discussed in the paper but the equation set emerging from the imposition of the no-slip boundary on the plane wall needs to be solved numerically. Another interesting extension would be the possibility to study the transient deformation and asymptotic shapes of drop induced by insoluble surfactant and also the effect of gravity. This is another case where it would be important to complement other numerical works present in the literature \cite{Dimi1,Dimi2,Li,Yon}.  Finally, all results presented in this paper are also useful in the context of ensemble averaging techniques to formulate effective boundary condition problems  \cite{Prosper1,Prosper2,Stone,JFM,KAZU}, an issue which has been recently renewed due to its relevance in small scale hydrodynamics \cite{Lauga}.

\section{Appendix}

\appendix\label{app}

In this appendix we report the recurrence formulae to determine the coefficients $A_{2n+1}$, $A_{2n}$, $G_{2n}$, $\hat{A}_{2n-1}$, $\hat{A}_{2n}$ and $\hat{G}_{2n-2}$ for $n\geq 1$. The relevant boundary conditions at $r=1$ are given by (\ref{eq:kinconou01})-(\ref{eq:stressbc02}). Using the general representations (\ref{eq:gen_out_w01})-(\ref{eq:gen_in_v01}), we can rewrite the boundary conditions and apply definite integrals for Legendre polynomials 
\be\nonumber
\int_0^{1}\!\!\!\!{\rm d}\mu\ 
P_{2n-1}^m(\mu)P_{2k-1}^m(\mu)=
\frac{\delta_{nk}}{(4k-1)}
\frac{(2k-1+m)!}{(2k-1-m)!},
\ee
\be\nonumber
\int_0^{1}\!\!\!\!{\rm d}\mu\ 
P_{2n}^m(\mu)P_{2k}^m(\mu)=
\frac{\delta_{nk}}{(4k+1)}
\frac{(2k+m)!}{(2k-m)!},
\ee
\be\nonumber
\int_0^{1}\!\!\!\!{\rm d}\mu\ 
P_{2n}(\mu)P_{2k-1}(\mu)=
\frac{(-1)^{n+k}(2n-1)!!(2k-1)!!}{2(2n-2k+1)(n+k)(2n)!!(2k-2)!!},
\ee
\be\nonumber
\int_0^{1}\!\!\!\!{\rm d}\mu\ 
P_{2n}^1(\mu)P_{2k-1}^1(\mu)=
\frac{(-1)^{n+k}(2n+1)!!(2k-1)!!}{2(2n-2k+1)(n+k)(2n-2)!!(2k-2)!!},
\ee
\be\nonumber
\int_0^{1}\!\!\!\!{\rm d}\mu\ 
P_{2n}^2(\mu)P_{2k-1}^2(\mu)=
\frac{(-1)^{n+k}(2n+1)!!(2k+1)!!}{2(2n-2k+1)(n+k)(2n-2)!!(2k-4)!!}.
\ee
In this way, we can eliminate the summation for the even-order mode and we finally obtain the recurrence formulae:
\begin{align}
\frac{\delta_{k1}}{3}=&
\frac{(2k+1)}{2(4k-1)}A_{2k}
+\frac{(k+1)(2k+3)(8k^2+6k-3)}{2k(2k+1)(4k+3)(4k+5)}A_{2k+2}
\nonumber\\&
+\frac{1}{2k(2k+1)}G_{2k+2}
-\Delta_{k}\sum_{n=1}^\infty
(n+1){\Lambda}_{n,k}A_{2n+1}\nonumber
\end{align}
\begin{align}
\frac{\delta_{k1}}{3}=&
\frac{(k-1)(2k-1)(4k^2+k-2)}{k(2k+1)(4k-3)(4k-1)}\hat{A}_{2k-2}
+\frac{k}{4k+3}\hat{A}_{2k}
\nonumber\\&
+\frac{1}{2k(2k+1)}\hat{G}_{2k-2}
-\Delta_k\sum_{n=1}^\infty
(2n-1)\hat{\Lambda}_{n,k}\hat{A}_{2n-1},\nonumber
\end{align}
\begin{align}
&
\frac{6k^2-7k-2}{(2k-1)(4k-1)(4k+1)}
A_{2k}
+\frac{(2k+3)(6k^2+5k-3)}{2k(2k+1)(4k+3)(4k+5)}A_{2k+2}\nonumber\\&
+\frac{1}{(2k-1)(4k-1)}G_{2k} 
+\frac{(2k+3)}{2k(2k+1)(4k+3)}G_{2k+2}\nonumber\\&
-\frac{\Delta_k}{(2k-1)(2k+2)}
\sum_{n=1}^\infty (n-1)(2n+3){\Lambda}_{n,k}A_{2n+1}\nonumber\\=&
-\frac{(k-1)(6k^2+k-4)}{k(2k+1)(4k-3)(4k-1)}\hat{A}_{2k-2}
-\frac{6k^2+13k+3}{2(k+1)(4k+1)(4k+3)}\hat{A}_{2k}\nonumber\\&
+\frac{k-1}{k(2k+1)(4k-1)}\hat{G}_{2k-2}
+\frac{1}{2(k+1)(4k+3)}\hat{G}_{2k}\nonumber\\&
+\frac{\Delta_k}{(k+1)(2k-1)}
\sum_{n=1}^\infty
(n-1)(2n+3)\hat{\Lambda}_{n,k}\hat{A}_{2n-1},\nonumber
\end{align}
\begin{align}
&
-\frac{2k^2(2k+1)}{(4k-1)(4k+1)}A_{2k}
-\frac{(k+1)(2k+1)(2k+3)}{(4k+3)(4k+5)}A_{2k+2} \nonumber\\&
+\frac{2k}{4k-1}G_{2k}+\frac{2k+1}{4k+3}G_{2k+2}
+\Delta_{k}\sum_{n=1}^\infty n(2n+1){\Lambda}_{n,k}A_{2n+1}
\nonumber\\=&
\frac{2k(k-1)(2k-1)}{(4k-3)(4k-1)}\hat{A}_{2k-2}
+\frac{k(2k+1)^2}{(4k+1)(4k+3)}\hat{A}_{2k}\nonumber\\&
+\frac{2k}{4k-1}\hat{G}_{2k-2}
+\frac{2k+1}{4k+3}\hat{G}_{2k}
-2\Delta_{k}
\sum_{n=1}^\infty
n(2n+1)\hat{\Lambda}_{n,k}\hat{A}_{2n-1},\nonumber
\end{align}
\begin{align}
&
-\frac{(2k+1)(6k^2-7k-2)}{(2k-1)(4k-1)(4k+1)}A_{2k}
-\frac{(2k+3)^2(6k^2+5k-3)}{2k(2k+1)(4k+3)(4k+5)}A_{2k+2}
\nonumber\\&
-\frac{(2k+1)}{(2k-1)(4k-1)}G_{2k}
-\frac{(2k+3)^2}{2k(2k+1)(4k+3)}G_{2k+2}
\nonumber\\&
+\frac{\Delta_k}{(k+1)(2k-1)}
\sum_{n=1}^\infty
(n-1)(n+1)(2n+3){\Lambda}_{n,k}A_{2n+1}
\nonumber\\=&
\lambda\Biggl\{
-\frac{2(k-1)^2(6k^2+k-4)}{k(2k+1)(4k-3)(4k-1)}
\hat{A}_{2k-2}
-\frac{k(6k^2+13k+3)}{(k+1)(4k+1)(4k+3)}
\hat{A}_{2k} \nonumber\\&
+\frac{2(k-1)^2}{k(2k+1)(4k-1)}\hat{G}_{2k-2}
+\frac{k}{(k+1)(4k+3)}
\hat{G}_{2k}
\Biggr\}
\nonumber\\&
+
\frac{\lambda\Delta_k}{(k+1)(2k-1)}
\sum_{n=1}^\infty
(n-1)(2n-1)(2n+3)\hat{\Lambda}_{n,k}\hat{A}_{2n-1},\nonumber
\end{align}
\begin{align}
&
\frac{2k^2(2k+1)^2}{(4k-1)(4k+1)}
A_{2k}
+\frac{(k+1)(2k+1)(2k+3)^2}{(4k+3)(4k+5)}
A_{2k+2}
\nonumber\\&
-\frac{2k(2k+1)}{(4k-1)}
G_{2k  }
-\frac{(2k+1)(2k+3)}{(4k+3)}
G_{2k+2}
\nonumber\\&
-2\Delta_k
\sum_{n=1}^\infty
n(n+1)(2n+1){\Lambda}_{n,k}A_{2n+1}
\nonumber\\=&
\lambda\Biggl\{
\frac{4k(k-1)^2(2k-1)}{(4k-3)(4k-1)}
\hat{A}_{2k-2}
+\frac{2k^2(2k+1)^2}{(4k+1)(4k+3)}
\hat{A}_{2k}
\nonumber\\&
+\frac{4k(k-1)}{4k-1}
\hat{G}_{2k-2}
+\frac{2k(2k+1)}{4k+3}
\hat{G}_{2k}
\Biggr\}
\nonumber\\&
-2\lambda\Delta_{k}
\sum_{n=1}^\infty
n(2n-1)(2n+1)
\hat{\Lambda}_{n,k}\hat{A}_{2n-1},\nonumber
\end{align}
where we have used
\begin{equation}
\begin{split}
\Lambda_{n,k}=&
\frac{(-1)^{n}(2n+1)!!}
{2(2n-2k-1)(2n-2k+1)(n+k)(n+k+1)(2n)!!},\\
\hat{\Lambda}_{n,k}=&
\frac{(-1)^{n}(2n-1)!!}
{4(2n-2k-1)(2n-2k+1)(n+k)(n+k+1)(2n-2)!!},\nonumber
\end{split}
\end{equation}
and 
\begin{equation}
\Delta_k=\frac{(-1)^{k}(4k+1)(2k-1)!!}{(2k)!!}.\nonumber
\end{equation}

\section{Acknowledgements}

We are  indebted to Prof. A. Prosperetti for proposing this problem and for his constant suggestions during the preparation of the manuscript. M. S. is grateful to B. M. Borkent for pertinent suggestions and STW (nanoned Programme) for financial support.

\newpage


\begin{center}
\begin{table}
\caption{Coefficients for $\lambda=0$.}
\label{tab:coef_chi0}
{\footnotesize
\begin{tabular}{rrrrrrr}
\hline
$n$
&\multicolumn{1}{c}{$A_{2n+1}$}
&\multicolumn{1}{c}{$A_{2n}$}
&\multicolumn{1}{c}{$G_{2n}$}
&\multicolumn{1}{c}{$\hat{A}_{2n-1}$}
&\multicolumn{1}{c}{$\hat{A}_{2n}$}
&\multicolumn{1}{c}{$\hat{G}_{2n-2}$}\\\hline
 1&$-4.2435\times 10^{-12}$&$ 0.66667              $&$ 0.40000              $&$-11.47649   $&$ 2.53882   $&$ 8.03099  $\\
 2&$-2.3846\times 10^{-12}$&$-9.0339\times 10^{-14}$&$-2.5928\times 10^{-13}$&$- 2.13519   $&$ 0.17201   $&$ 5.42824  $\\
 3&$ 1.1579\times 10^{-12}$&$-1.8727\times 10^{-14}$&$ 2.8944\times 10^{-13}$&$  0.24983   $&$-0.071796  $&$ 0.12475  $\\
 4&$-1.8929\times 10^{-12}$&$-7.2444\times 10^{-14}$&$-1.8123\times 10^{-13}$&$- 0.062825  $&$ 0.036175  $&$-0.082789 $\\
 5&$-2.1743\times 10^{-13}$&$ 1.6780\times 10^{-14}$&$ 2.5391\times 10^{-13}$&$  0.019726  $&$-0.020377  $&$ 0.056309 $\\
 6&$-2.7782\times 10^{-13}$&$-1.5622\times 10^{-14}$&$-2.1449\times 10^{-13}$&$- 0.0063530 $&$ 0.012379  $&$-0.039535 $\\
 7&$ 3.3070\times 10^{-15}$&$ 1.3030\times 10^{-14}$&$ 1.6910\times 10^{-13}$&$  0.0015681 $&$-0.0079442 $&$ 0.028544 $\\
 8&$-8.3398\times 10^{-14}$&$ 4.3793\times 10^{-15}$&$-2.0627\times 10^{-13}$&$  0.00024267$&$ 0.0053167 $&$-0.021091 $\\
 9&$-7.9357\times 10^{-15}$&$ 3.6435\times 10^{-14}$&$ 9.0663\times 10^{-14}$&$- 0.00090699$&$-0.0036784 $&$ 0.015882 $\\
10&$-1.8344\times 10^{-13}$&$-1.1164\times 10^{-14}$&$-1.4936\times 10^{-13}$&$  0.0011024 $&$ 0.0026142 $&$-0.012147 $\\
11&$ 1.7744\times 10^{-14}$&$ 1.6463\times 10^{-14}$&$ 1.3542\times 10^{-13}$&$- 0.0011030 $&$-0.0018995 $&$ 0.0094079$\\
12&$-3.0430\times 10^{-13}$&$ 6.7278\times 10^{-15}$&$-2.5442\times 10^{-13}$&$  0.0010260 $&$ 0.0014061 $&$-0.0073615$\\
13&$-2.2858\times 10^{-13}$&$ 1.5070\times 10^{-14}$&$ 7.3790\times 10^{-14}$&$- 0.00092290$&$-0.0010573 $&$ 0.0058075$\\
14&$-2.6355\times 10^{-13}$&$-1.8224\times 10^{-14}$&$-5.9443\times 10^{-14}$&$  0.00081663$&$ 0.00080565$&$-0.0046107$\\
15&$ 2.3231\times 10^{-13}$&$-1.0819\times 10^{-14}$&$ 2.0463\times 10^{-13}$&$- 0.00071683$&$-0.00062094$&$ 0.0036777$\\
\hline
\end{tabular}
}
\end{table}
\end{center}

\begin{center}

\begin{table}
\caption{Coefficients for $\lambda=0.1$.}
\label{tab:coef_chi0.1}
{\footnotesize

\begin{tabular}{rrrrrrr}
\hline
$n$
&\multicolumn{1}{c}{$A_{2n+1}$}
&\multicolumn{1}{c}{$A_{2n}$}
&\multicolumn{1}{c}{$G_{2n}$}
&\multicolumn{1}{c}{$\hat{A}_{2n-1}$}
&\multicolumn{1}{c}{$\hat{A}_{2n}$}
&\multicolumn{1}{c}{$\hat{G}_{2n-2}$}\\\hline
 1&$-0.17436    $&$ 0.84972    $&$ 0.50983   $&$-10.46143  $&$ 2.31835   $&$ 7.49147  $\\
 2&$-0.077633   $&$ 0.15800    $&$-0.077149  $&$-1.94082   $&$ 0.15400   $&$ 4.92646  $\\
 3&$ 0.012187   $&$ 0.0094929  $&$-0.0011555 $&$ 0.22749   $&$-0.064461  $&$ 0.11746  $\\
 4&$-0.0035702  $&$-0.0046015  $&$ 0.0023703 $&$-0.057379  $&$ 0.032545  $&$-0.077592 $\\
 5&$ 0.0012341  $&$ 0.0025121  $&$-0.0022646 $&$ 0.018100  $&$-0.018362  $&$ 0.052686 $\\
 6&$-0.00042666 $&$-0.0014877  $&$ 0.0019020 $&$-0.0058827 $&$ 0.011167  $&$-0.036976 $\\
 7&$ 0.00011347 $&$ 0.00093504 $&$-0.0015382 $&$ 0.0014964 $&$-0.0071731 $&$ 0.026705 $\\
 8&$ 0.000013522$&$-0.00061510 $&$ 0.0012302 $&$ 0.00017241$&$ 0.0048045 $&$-0.019748 $\\
 9&$-0.000063680$&$ 0.00041948 $&$-0.00098250$&$-0.00079079$&$-0.0033262 $&$ 0.014888 $\\
10&$ 0.000080443$&$-0.00029453 $&$ 0.00078655$&$ 0.00097802$&$ 0.0023654 $&$-0.011407 $\\
11&$-0.000082443$&$ 0.00021183 $&$-0.00063199$&$-0.00098504$&$-0.0017197 $&$ 0.0088482$\\
12&$ 0.000078085$&$-0.00015543 $&$ 0.00050977$&$ 0.00091948$&$ 0.0012736 $&$-0.0069382$\\
13&$-0.000071276$&$ 0.00011601 $&$-0.00041261$&$-0.00082899$&$-0.00095812$&$ 0.0054870$\\
14&$ 0.000063845$&$-0.000087837$&$ 0.00033493$&$ 0.00073472$&$ 0.00073043$&$-0.0043687$\\
15&$-0.000056630$&$ 0.000067335$&$-0.00027242$&$-0.00064571$&$-0.00056322$&$ 0.0034961$\\
\hline
\end{tabular}
}
\end{table}
\end{center}

\vspace{2em}

\begin{center}

\begin{table}
\caption{Coefficients for $\lambda=1$.}
\label{tab:coef_chi1}
{\footnotesize

\begin{tabular}{rrrrrrr}
\hline
$n$
&\multicolumn{1}{c}{$A_{2n+1}$}
&\multicolumn{1}{c}{$A_{2n}$}
&\multicolumn{1}{c}{$G_{2n}$}
&\multicolumn{1}{c}{$\hat{A}_{2n-1}$}
&\multicolumn{1}{c}{$\hat{A}_{2n}$}
&\multicolumn{1}{c}{$\hat{G}_{2n-2}$}\\\hline
 1&$-0.97542    $&$ 1.68344   $&$ 1.01007  $&$-5.85249    $&$ 1.31294   $&$ 5.05033  $\\
 2&$-0.42739    $&$ 0.87935   $&$-0.40999  $&$-1.06849    $&$ 0.077130  $&$ 2.67984  $\\
 3&$ 0.067950   $&$ 0.048954  $&$-0.014276 $&$ 0.12684    $&$-0.033073  $&$ 0.077672 $\\
 4&$-0.020266   $&$-0.024073  $&$ 0.017863 $&$-0.032570   $&$ 0.016955  $&$-0.049837 $\\
 5&$ 0.0071884  $&$ 0.013276  $&$-0.015674 $&$ 0.010543   $&$-0.0096669 $&$ 0.033457 $\\
 6&$-0.0026054  $&$-0.0079212 $&$ 0.012751 $&$-0.0035922  $&$ 0.0059251 $&$-0.023387 $\\
 7&$ 0.00079580 $&$ 0.0050080 $&$-0.010168 $&$ 0.0010494  $&$-0.0038295 $&$ 0.016889 $\\
 8&$-0.000042750$&$-0.0033104 $&$ 0.0080860$&$-0.000054498$&$ 0.0025779 $&$-0.012520 $\\
 9&$-0.00026832 $&$ 0.0022669 $&$-0.0064524$&$-0.00033322 $&$-0.0017925 $&$ 0.0094793$\\
10&$ 0.00038388 $&$-0.0015974 $&$ 0.0051775$&$ 0.00046672 $&$ 0.0012795 $&$-0.0073022$\\
11&$-0.00041078 $&$ 0.0011527 $&$-0.0041798$&$-0.00049081 $&$-0.00093345$&$ 0.0057059$\\
12&$ 0.00039788 $&$-0.00084839$&$ 0.0033941$&$ 0.00046853 $&$ 0.00069351$&$-0.0045116$\\
13&$-0.00036830 $&$ 0.00063500$&$-0.0027709$&$-0.00042837 $&$-0.00052328$&$ 0.0036025$\\
14&$ 0.00033312 $&$-0.00048214$&$ 0.0022727$&$ 0.00038336 $&$ 0.00040007$&$-0.0029001$\\
15&$-0.00029762 $&$ 0.00037060$&$-0.0018715$&$-0.00033936 $&$-0.00030935$&$ 0.0023503$\\
\hline
\end{tabular}
}
\end{table}
\end{center}

\begin{center}

\begin{table}
\caption{Coefficients for $\lambda=10$.}
\label{tab:coef_chi10}
{\footnotesize

\begin{tabular}{rrrrrrr}
\hline
$n$
&\multicolumn{1}{c}{$A_{2n+1}$}
&\multicolumn{1}{c}{$A_{2n}$}
&\multicolumn{1}{c}{$G_{2n}$}
&\multicolumn{1}{c}{$\hat{A}_{2n-1}$}
&\multicolumn{1}{c}{$\hat{A}_{2n}$}
&\multicolumn{1}{c}{$\hat{G}_{2n-2}$}\\\hline
 1&$-1.83516   $&$ 2.55162   $&$ 1.53097  $&$-1.10110    $&$ 0.25418    $&$ 2.56548   $\\
 2&$-0.78449   $&$ 1.64184   $&$-0.68804  $&$-0.19612    $&$ 0.011796   $&$ 0.47557   $\\
 3&$ 0.12819   $&$ 0.079556  $&$-0.050045 $&$ 0.023929   $&$-0.0054338  $&$ 0.018141  $\\
 4&$-0.039327  $&$-0.040886  $&$ 0.045731 $&$-0.0063205  $&$ 0.0028862  $&$-0.011069  $\\
 5&$ 0.014454  $&$ 0.023098  $&$-0.036923 $&$ 0.0021199  $&$-0.0016810  $&$ 0.0072831 $\\
 6&$-0.0055521 $&$-0.013997  $&$ 0.029024 $&$-0.00076551 $&$ 0.0010455  $&$-0.0050478 $\\
 7&$ 0.0019496 $&$ 0.0089486 $&$-0.022775 $&$ 0.00025708 $&$-0.00068302 $&$ 0.0036351 $\\
 8&$-0.00039965$&$-0.0059665 $&$ 0.017979 $&$-0.000050954$&$ 0.00046375 $&$-0.0026961 $\\
 9&$-0.00027530$&$ 0.0041147 $&$-0.014313 $&$-0.000034189$&$-0.00032475 $&$ 0.0020468 $\\
10&$ 0.00055414$&$-0.0029172 $&$ 0.011495 $&$ 0.000067373$&$ 0.00023324 $&$-0.0015835 $\\
11&$-0.00064763$&$ 0.0021163 $&$-0.0093099$&$-0.000077380$&$-0.00017109 $&$ 0.0012443 $\\
12&$ 0.00065388$&$-0.0015652 $&$ 0.0075980$&$ 0.000076998$&$ 0.00012775 $&$-0.00099055$\\
13&$-0.00062032$&$ 0.0011769 $&$-0.0062438$&$-0.000072149$&$-0.000096843$&$ 0.00079718$\\
14&$ 0.00057041$&$-0.00089748$&$ 0.0051623$&$ 0.000065643$&$ 0.000074373$&$-0.00064750$\\
15&$-0.00051578$&$ 0.00069277$&$-0.0042910$&$-0.000058812$&$-0.000057759$&$ 0.00053005$\\
\hline
\end{tabular}
}
\end{table}
\end{center}

\begin{center}

\begin{table}
\caption{Coefficients for $\lambda=10^{100}$.}
\label{tab:coef_chi1e100}
{\footnotesize

\begin{tabular}{rrrrrrr}
\hline
$n$
&\multicolumn{1}{c}{$A_{2n+1}$}
&\multicolumn{1}{c}{$A_{2n}$}
&\multicolumn{1}{c}{$G_{2n}$}
&\multicolumn{1}{c}{$\hat{A}_{2n-1}$}
&\multicolumn{1}{c}{$\hat{A}_{2n}$}
&\multicolumn{1}{c}{$\hat{G}_{2n-2}$}\\\hline
 1&$-2.04530   $&$ 2.75543   $&$ 1.65326  $&$-1.2272\times 10^{-99} $&$ 2.8662\times 10^{-100}$&$ 2.00000	       $\\
 2&$-0.86783   $&$ 1.82575   $&$-0.73104  $&$-2.1696\times 10^{-100}$&$ 1.2213\times 10^{-101}$&$ 5.1794\times 10^{-100}$\\
 3&$ 0.14321   $&$ 0.084338  $&$-0.063687 $&$ 2.6733\times 10^{-101}$&$-5.8109\times 10^{-102}$&$ 2.1417\times 10^{-101}$\\
 4&$-0.044300  $&$-0.044198  $&$ 0.054743 $&$-7.1196\times 10^{-102}$&$ 3.1280\times 10^{-102}$&$-1.2864\times 10^{-101}$\\
 5&$ 0.016436  $&$ 0.025196  $&$-0.043364 $&$ 2.4106\times 10^{-102}$&$-1.8352\times 10^{-102}$&$ 8.4124\times 10^{-102}$\\
 6&$-0.0064047 $&$-0.015350  $&$ 0.033811 $&$-8.8306\times 10^{-103}$&$ 1.1467\times 10^{-102}$&$-5.8150\times 10^{-102}$\\
 7&$ 0.0023185 $&$ 0.0098491 $&$-0.026427 $&$ 3.0574\times 10^{-103}$&$-7.5171\times 10^{-103}$&$ 4.1832\times 10^{-102}$\\
 8&$-0.00054568$&$-0.0065846 $&$ 0.020824 $&$-6.9572\times 10^{-104}$&$ 5.1171\times 10^{-103}$&$-3.1021\times 10^{-102}$\\
 9&$-0.00023620$&$ 0.0045508 $&$-0.016566 $&$-2.9333\times 10^{-104}$&$-3.5909\times 10^{-103}$&$ 2.3560\times 10^{-102}$\\
10&$ 0.00056676$&$-0.0032322 $&$ 0.013305 $&$ 6.8908\times 10^{-104}$&$ 2.5837\times 10^{-103}$&$-1.8242\times 10^{-102}$\\
11&$-0.00068450$&$ 0.0023485 $&$-0.010782 $&$-8.1786\times 10^{-104}$&$-1.8982\times 10^{-103}$&$ 1.4350\times 10^{-102}$\\
12&$ 0.00070095$&$-0.0017395 $&$ 0.0088074$&$ 8.2541\times 10^{-104}$&$ 1.4194\times 10^{-103}$&$-1.1439\times 10^{-102}$\\
13&$-0.00067033$&$ 0.0013096 $&$-0.0072468$&$-7.7965\times 10^{-104}$&$-1.0775\times 10^{-103}$&$ 9.2200\times 10^{-103}$\\
14&$ 0.00061962$&$-0.00099993$&$ 0.0060008$&$ 7.1306\times 10^{-104}$&$ 8.2854\times 10^{-104}$&$-7.5021\times 10^{-103}$\\
15&$-0.00056237$&$ 0.00077279$&$-0.0049970$&$-6.4124\times 10^{-104}$&$-6.4425\times 10^{-104}$&$ 6.1534\times 10^{-103}$\\
\hline
\end{tabular}
}
\end{table}
\end{center}

\end{document}